# A Zonal Similarity Analysis Of Friction Factors: Case Study 1: Pipe Flow Of Power Law Fluids

**Trinh, Khanh Tuoc**

Institute Of Food Nutrition And Human Health

Massey University, New Zealand

K.T.Trinh@massey.ac.nz

## Abstract

A zonal similarity analysis of friction factors in pipe flow of power law fluids is presented. It uses the critical Reynolds number and friction factor at transition as estimates of the normalised velocity and distance between the wall and log-law layers and successfully collapses both Newtonian and non-Newtonian data for laminar, transition and turbulent flows.

## 1  Introduction

Friction losses in pipe flow of non-Newtonian fluids have been investigated for almost a century. The most popular representation shows the friction factor

$$f = \frac{2\tau_w}{\rho V^2} \quad (1)$$

against the so-called against the so-called generalised Metzner–Reed (1955) number

$$\mathrm{Re}_g = \frac{D^{n'} V^{2-n'} \rho}{K' 8^{n'-1} \left(\frac{3n'+1}{4n'}\right)^{n'}} \quad (2)$$

Where $V$ is the average velocity, $\tau_w$ the wall shear stress, $\rho$ the fluid density and the rheological parameters $K'$ and $n'$ are determined from viscometric flow data with the equation

$$\tau_w = K' \left(\frac{8V}{D}\right)^{n'} \quad (3)$$

Turbulent flow data falls in a family of curves with parameter $n'$ as shown in Figure 1. An extensive review of previous correlations has been presented elsewhere (Trinh, 2009a). In the same paper, it was been shown that all the data in Figure 1 collapse into a unique master curve when the friction factor and Reynolds number are expressed in term of the instantaneous wall shear stress at the point of bursting instead of the time-averaged wall shear stress.

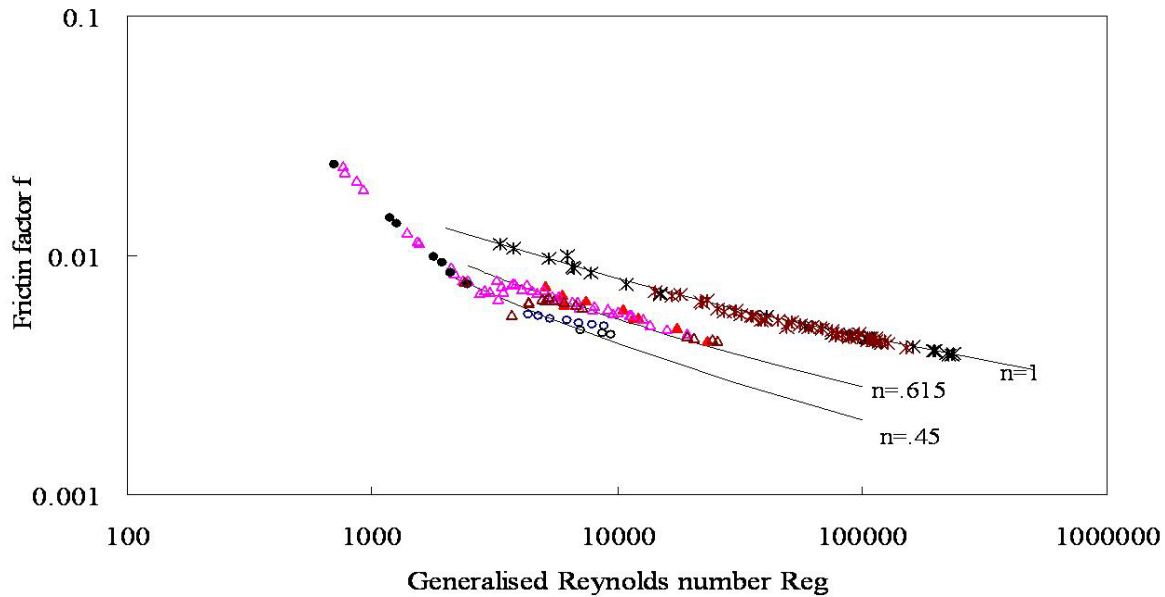

Figure 1. Plot of friction factor vs. Metzner-Reed Reynolds number for 3 values of $n'$. Data of Dodge, Bogue and Yoo

While a unique master instantaneous friction factor curve is helpful in showing the universality of the turbulence mechanism in all fluids – in contrast to claims of separate mechanisms for non-Newtonian fluids (Trinh, 2009a, Trinh, 2010d) – practical applications normally require the estimate of the time averaged friction factor. This can easily be achieved with the enhanced insight for specific fluid models e.g. (Trinh, 2010b) but a master curve applicable to a large spectrum of fluid models and flow geometries will be more convenient. It is the subject of this paper.

Obot (1993) is the only other author that I have found to attempt this task. He argued heuristically that the conditions at transition must somehow define the structure of the subsequent turbulent flows. The critical transition Reynolds number and friction factor can be used as a scaling factor for flow and heat transfer in channels of different geometries, two phase flow and mixing. He called this similarity plot between $\text{Re}_m = \text{Re}/\text{Re}_c$ and $f_m = f/f_c$, where $\text{Re}_c$ is the critical Reynolds number at transition, "the frictional law of corresponding states". Malosova et al. (2006) have also produced a similarity curve for Kaolin suspensions by plotting $f\,\text{Re}_m$ against $\text{Re}_m$ but did not discuss the principles behind the method.

## 2    Theory

In a previous paper (Trinh, 2010a), it was shown that the velocity profiles of turbulent flows can be divided into three distinct layers: a wall layer, a log-law layer and a law-of-the-wake layer that are not mutually destructive but additive. The wall layer has the same velocity profile as a laminar boundary layer and the two other layers are created by the intrusion of ejected wall fluids into the region outside the wall layer. Thus the three layers are governed by different mechanisms and it is not possible to obtain a universal velocity profile by normalising the local velocity $U$ and distance $y$ with the wall parameters, the kinematic viscosity $\nu$ and the friction velocity $u_* = \sqrt{\tau_w/\rho}$ where $\tau_w$ is the wall shear stress. Normalising with the values $U_\nu$ and $y = \delta_\nu$ at the interface between the wall and log-law layers does produce a unique zonal similarity curve for the wall and log-law layers in flows of all fluids in all geometries (Trinh, 2010d) as shown in Figure 1.

The velocity profile in the law-of-the wake layer is a function of the pressure distribution and the flow geometry as shown in Figure 1. For channel flow, this layer is relatively thin and the friction factor is usually calculated by fitting the whole field with only a wall layer and a log-law ((Prandtl, 1935, Karman, 1934). Using this well-known approximation we conclude from the zonal similarity analysis that there is a unique relation between the normalised maximum velocity $U_m/U\nu$ and radius $R/\delta_\nu$

$$\frac{U_m^+}{U_\nu^+} = f\left(\frac{R^+}{\delta_\nu^+}\right) \tag{4}$$

The normalising parameters $U_\nu^+$ and $\delta_\nu^+$ are not available in most experimental measurements of the friction factor but can be taken as the values of $\left(U_m^+, R^+\right)$ at the critical transition Reynolds number (Trinh, 2010c).

Since

$$V^+ = \sqrt{\frac{2}{f}} \tag{5}$$

Equation (1) may be rearranged as

$$\frac{\phi V^+}{\phi_c V_c^+} = \sqrt{\frac{\phi^2 f_c}{\phi_c^2 f}} = f\left(\frac{R^+}{R_c^+}\right) \tag{6}$$

where

$$\phi = \frac{U_m}{V} \tag{7}$$

As

$$R^+ = \frac{R u_*}{\nu} = \frac{RV}{\nu}\frac{u_*}{V} = \frac{\mathrm{Re}}{2}\sqrt{\frac{f}{2}} \tag{8}$$

we may further write

$$\left(\frac{\phi_c^2 f}{\phi^2 f_c}\right) = f'\left(\frac{\mathrm{Re}}{\mathrm{Re}_c}\right) \tag{9}$$

which points to a unique friction factor zonal similarity curve.

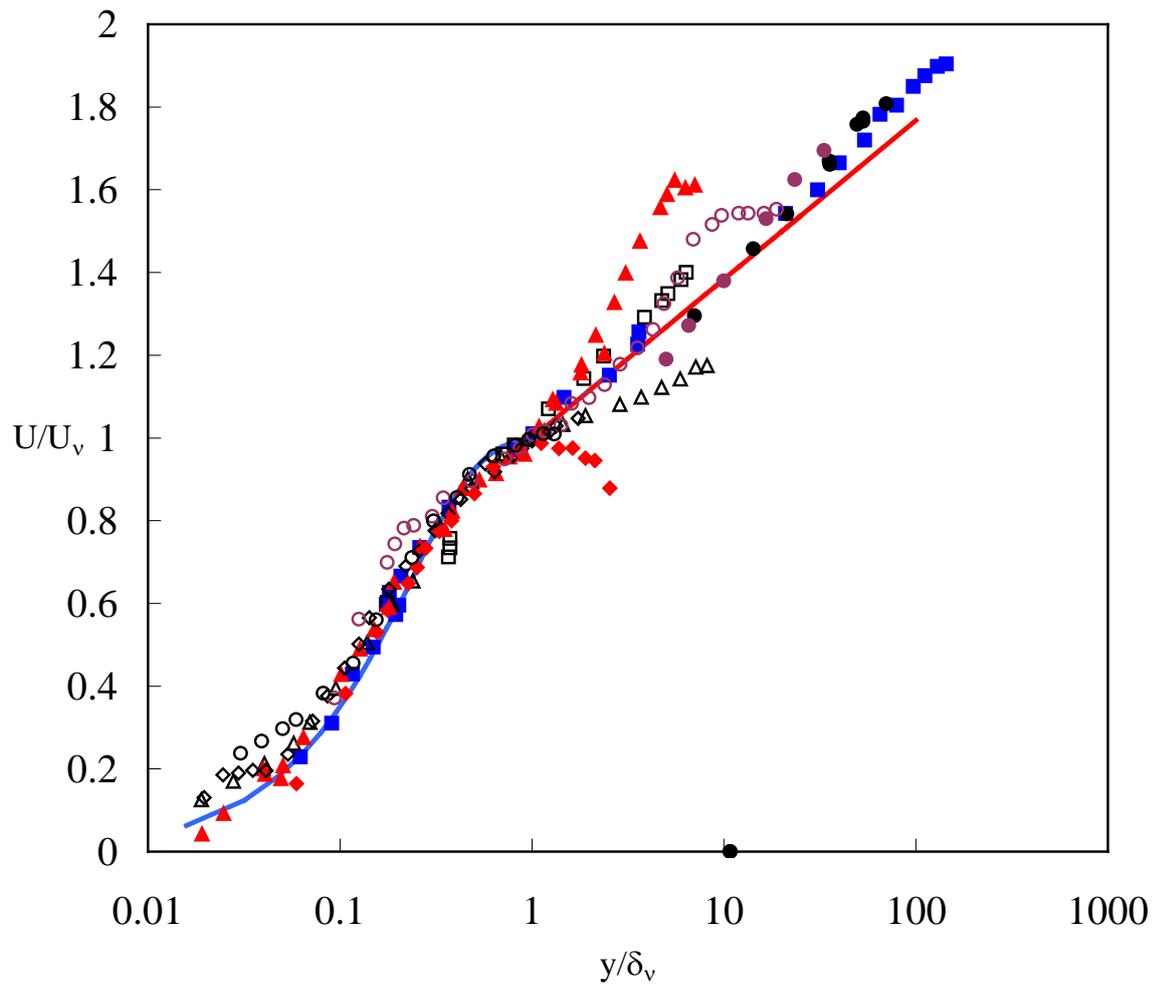

- ■ 1 Pipe flow Re = 500,000
- ● 2 Pipe flow, Re = 196000
- □ 3 Parallel plates, Re = 52300
- △ 4 Flat plate, dP/dx <<0
- ▲ 5 Flat plate, dP/dx >> 0
- ◇ 6 Converging channel, x = 500mm
- ◆ 7 Backward facing step, x/R = 0.48
- ○ 8 Wall riblets
- ○ 9 Viscoelastic flow, Re:167,000, n:0.64
- ● 10 Power law fluid, Re:17,500, n:465
- —— 11 Stokes averaged velocity profile
- —— 12 log-law

Figure 2. Zonal similar velocity profile. From Trinh (2010b)

# 3 Comparison with published literature data

The argument is tested with experimental data from the theses of Dodge (1959), Bogue (1962) and Yoo (1974). The critical transition pipe Reynolds number has been shown (Trinh, 2010c) to be

$$\text{Re}_{g,c} = 2010\left(\frac{3n+1}{4n}\right) \tag{10}$$

The critical transition friction factor is

$$f_c = \frac{16}{2010\left(\frac{3n+1}{4n}\right)} \tag{11}$$

The velocity ratio is given by

$$\phi = \frac{U_m^+}{V^+} = \left(1 - \Delta U^+\right)V^+ = \left(1 - \Delta U^+\sqrt{\frac{2}{f}}\right) \tag{12}$$

where

$$\Delta U^+ = U_m^+ - V^+ \tag{13}$$

The velocity difference $\Delta U^+$ is calculated for each value of set on $(n, R^+)$ by numerical integration of the velocity profile which is divided in two section: the wall layer given by the time averaged error function (Trinh, 2010d) and outer layer described approximated by the logarithmic law of the wall. As a first approximation the law of the wake in channel flow is neglected (Trinh, 2010b).

Obot (op.cit.) never applied his friction law of corresponding states to non-Newtonian fluids. A plot for power law fluids is shown in Figure 3.

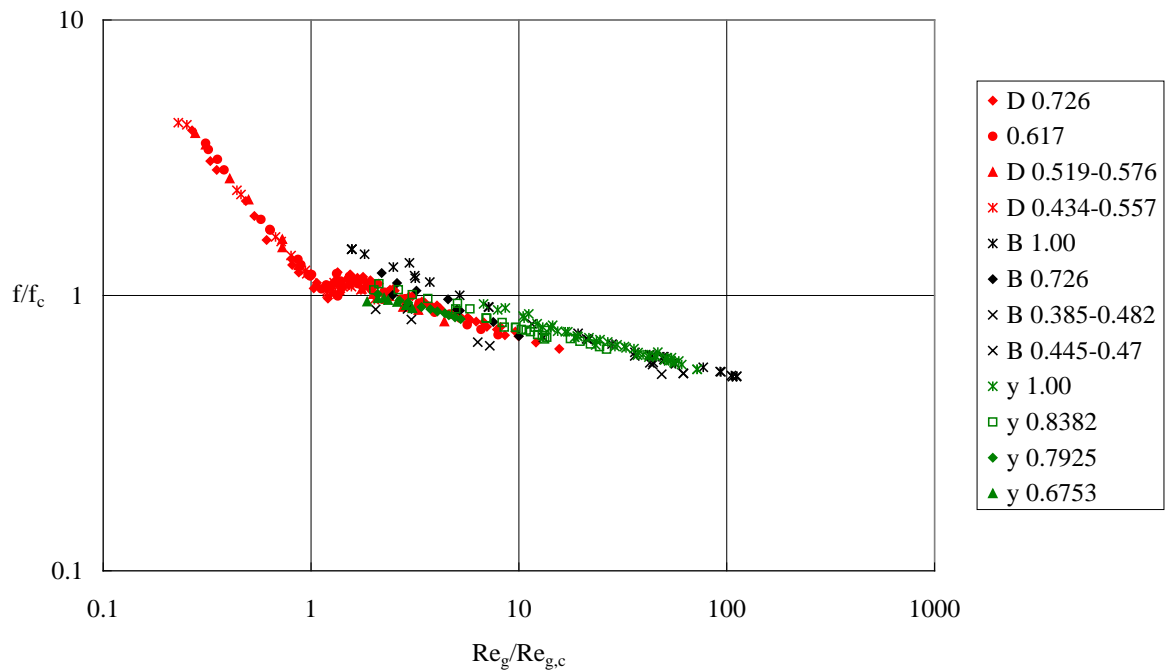

Figure 3  Plot of friction factor and Reynolds number ratio. Numbers indicate the range of $n$ values. Data source (red D) Dodge, (black B) Bogue, (green Y) Yoo.

The zonal similarity curve for power law pipe friction factors is shown in Figure 4.

## 4      Discussion

Figures 3 and 4 demonstrate the possibility of collapsing time-averaged friction factors of non-Newtonian fluids onto a single master curve that can be prepared from much more extensive and reliable measurements of Newtonian friction factors. The method also collapses data from other channel geometries as shown by Obot (1993) except when form drag is involved.

There is however a clear difference between the underlying physical visualisation in the present work  and that of Obot. Obot believed that the critical transition Reynolds number somehow set up the state of turbulence at higher Reynolds numbers in the same physical system but did not elaborate on the details of that interrelationship.

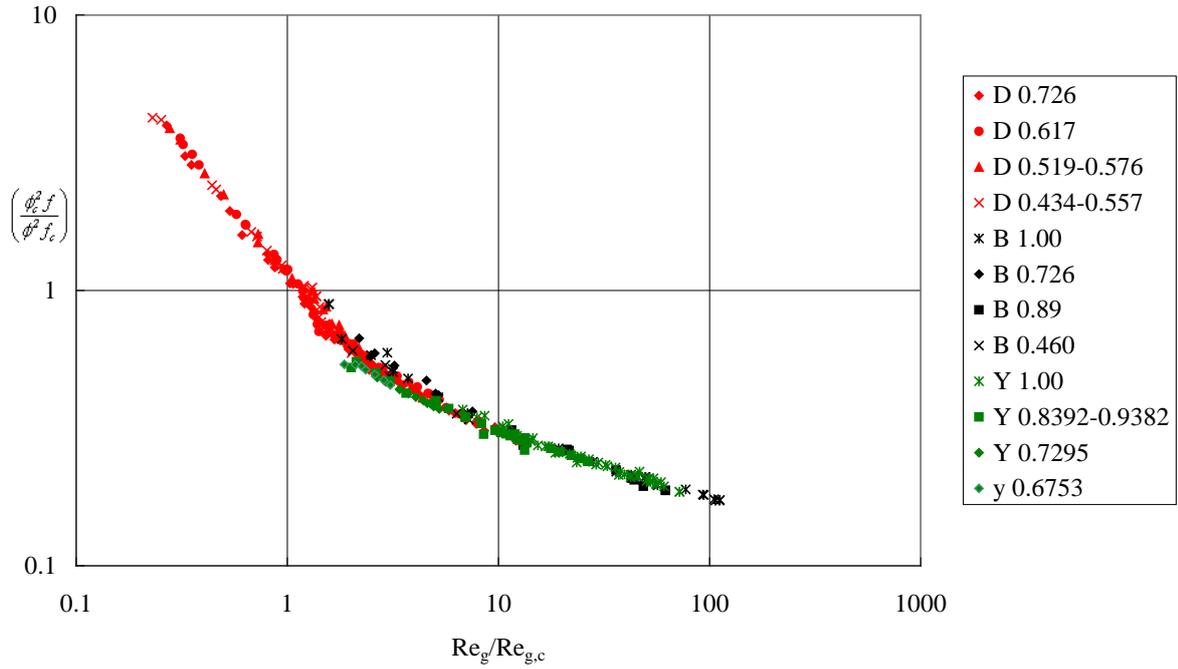

Figure 4 Zonal similarity friction factor plot. Numbers indicate the range of $n$ values. Data source (red D) Dodge, (black B) Bogue, (green Y) Yoo.

The present approach starts from the normalisation of the velocity profile with the velocity and distance scale at the interface between the wall and log-law layers $U_v, \delta_v$. The critical Reynolds number and friction factor $\text{Re}_c, f_c$ are simply estimates of $U_v, \delta_v$ that are not readily available in practical applications. The asymptotic values of $U_v, \delta_v$ are reached at the end of the laminar regime and appear to be the same for channel flows of many geometries (Trinh, 2010c) and we should not be surprised that the same master curve applies to all the geometries shown in Obot's paper (Obot, 1993). The point of transition is best correlated with the parameter

$$\varepsilon = \frac{U}{D\omega} = \left(\frac{UD}{\nu}\right)\left(\frac{\nu}{D^2\omega}\right) \qquad (14)$$

which includes the Reynolds number but also the Strouhal number of the disturbances introduced typical by entrance conditions (Trinh, 2009b). If we use $\varepsilon_c$ as a the criterion for transition, then $\text{Re}_c$ will vary with the entrance design that affect $\nu/D^2\omega$ as shown by the data of Tam and Ghajar in Figure 5.

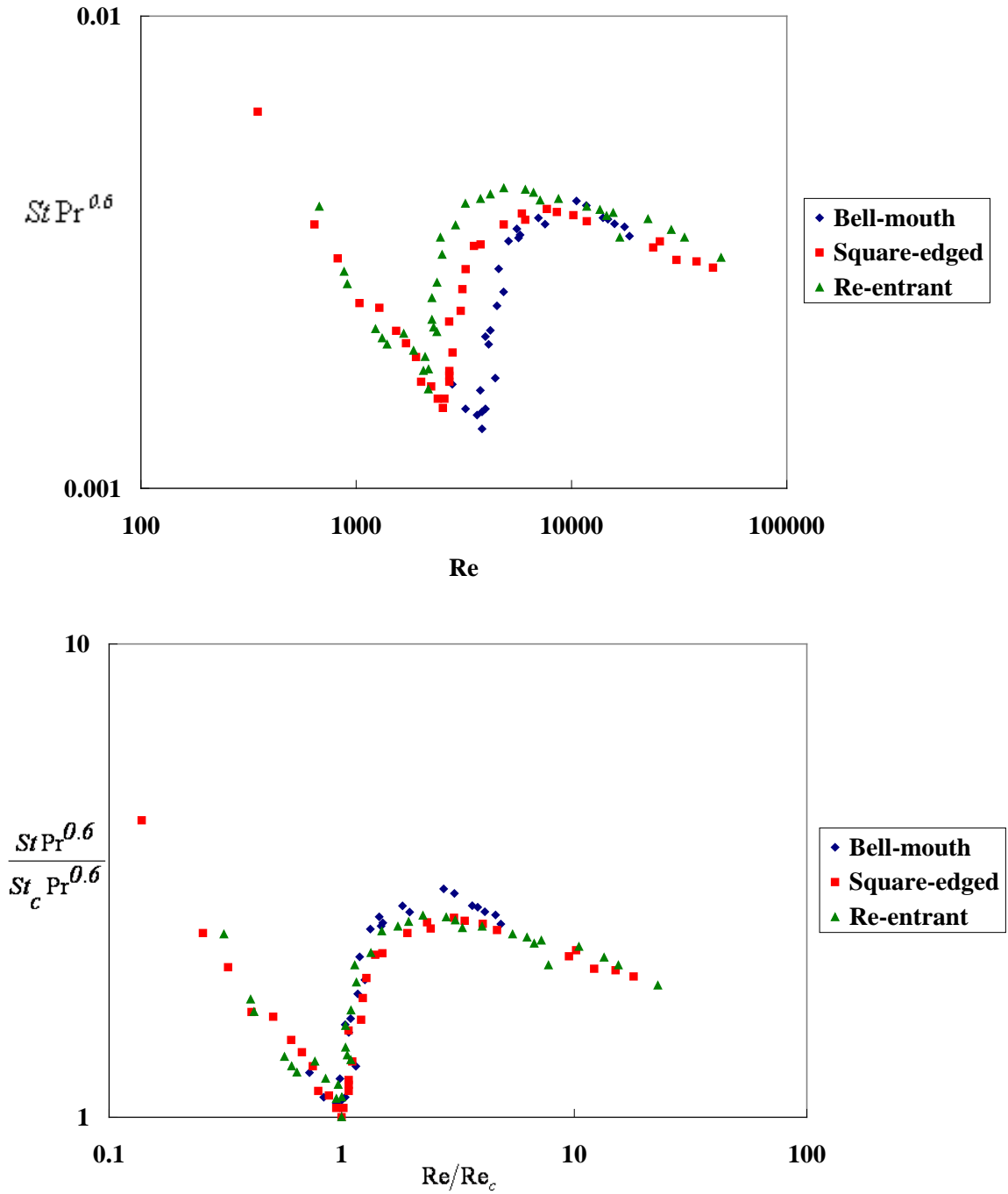

Figure 5  Effect of entrance design on heat transfer. Data (Tam and Ghajar, 1994) Top Stanton vs. Reynolds number. Bottom Normalised data

Normalising the Stanton and Reynolds numbers with their critical transition values again gives a unique similarity curve. The application of the method to pipe flow of power law fluids in this case study is reasonably simple. In fact many of the fluids studied by the authors quoted do not quite flow the power law but the log-log plot of $\tau_w$ vs. $8V/D$ of Carbopol for example is a curve, It has even be reported that Carbopol solutions can exhibit a yield stress (Zhu et al., 2005). In these situations I have adopted the approach of Dodge and Metzner (1959) also followed by Bogue and Yoo who treat the curve as sectionally linear near the friction factor measured. When we attempt to apply the technique with models with a yield stress like the Bingham Plastic or Herschel-Bulkley models the ratio $c = \tau_w/\tau_y$ which affects the apparent viscosity is a function of both the Reynolds number and the wall shear stress and introduce added difficulties. The data for these situations are being processed for further case studies.

The use of $\phi_c f/\phi f_c$ in Figure 4 instead of $f/f_c$ in Figure 3 does give a small decrease in the scatter of the data, especially in the transition region. It also changes the shape of the curve which shows in Figure 3 an increase in the ratio $f/f_c$ right at $\text{Re}_c$ before it starts to decrease again when full turbulence is achieved. In figure 4 $\phi_c f/\phi f_c$ decreases smoothly with increasing $\text{Re}/\text{Re}_c$. In further case studies it will be shown that the effect of the factor $\phi_c St \Pr^a/\phi St_c \Pr^a$ in the transition region is even more dramatic be cause the ratio $\phi = T_{\max}/T_{average}$ is a function not only of the Reynolds number but also of the Prandtl number. At the moment, there is no well accepted correlation for the transition region except perhaps for Gnielinski's for heat transfer in Newtonian pipe flow (1995). A master curve with a sharp representation for this region will be particularly useful in practical applications because many non-Newtonian fluids such as slurries, sauces,, thickeners and concentrated protein solutions are very viscous and processing equipment often operate in the transition regime.

## 5    Conclusion

It has been shown that the friction factor data for pipe flow of power law fluids can be collapsed into a unique zonal similarity curve for laminar, transition and turbulent flow.

## 6    References

BOGUE, D. C. 1962. *Velocity profiles in turbulent non-Newtonian pipe flow, Ph.D. Thesis,* University of Delaware.

DODGE, D. W. 1959. *Turbulent flow of non-Newtonian fluids in smooth round tubes, Ph.D. Thesis,* US, University of Delaware.
DODGE, D. W. & METZNER, A. B. 1959. Turbulent Flow of Non-Newtonian Systems. *AICHE Journal,* 5**,** 189-204.
GNIELINSKI, V. 1995. New Method To Calculate Heat-Transfer In The Transition Region Between Laminar And Turbulent Tube Flow. *Forschung Im Ingenieurwesen-Engineering Research,* 61**,** 240-&.
KARMAN, V. T. 1934. Turbulence and skin friction. *J. Aeronaut. Sci.,* 1**,** 1-20.
MALASOVA, I., MALKIN, A. Y., KHARATIYAN, E. & HALDENWANG, R. 2006. Scaling in pipeline flow of Kaolin suspensions. *J. Non-Newtonian Fluid Mech.,* 136**,** 76-78.
METZNER, A. B. & REED, J. C. 1955. Flow of Non-Newtonian Fluids - Correlation of the Laminar, Transition, and Turbulent-Flow Regions. *Aiche Journal,* 1**,** 434-440.
OBOT, N. T. 1993. The frictional law of correspondiong states: its origin and applications. *Trans IChemE,* 71 Part A**,** 3-10.
PRANDTL, L. 1935. The Mechanics of Viscous Fluids. *In:* W.F, D. (ed.) *Aerodynamic Theory III.* Berlin: Springer.
TAM, L. & GHAJAR, A. J. 1994. Heat Transfer Measurements and Correlations in the Transition Region for a Circular Tube with Three Different Inlet Configurations. *Experimental Thermal and Fluid Science,* 8**,** 79-90.
TRINH, K. T. 2009a. The Instantaneous Wall Viscosity in Pipe Flow of Power Law Fluids: Case Study for a Theory of Turbulence in Time-Independent Non-Newtonian Fluids. *arXiv.org 0912.5249v1 [phys.fluid-dyn]* [Online].
TRINH, K. T. 2009b. A Theory Of Turbulence Part I: Towards Solutions Of The Navier-Stokes Equations. *arXiv.org 0910.2072v1 [physics.flu.dyn.]* [Online].
TRINH, K. T. 2010a. Additive Layers: An Alternate Classification Of Flow Regimes. *arXiv.org 1001.1587 {phys.fluid-dyn}* [Online].
TRINH, K. T. 2010b. Logarithmic Correlations For Turbulent Pipe Flow Of Power Law Fluids. *arXiv.org [phys.fluid-dyn]* [Online]. Available: http://arxiv.org/abs/1007.0789.
TRINH, K. T. 2010c. On The Critical Reynolds Number For Transition From Laminar To Turbulent Flow. *arXiv.org [phys.fluid-dyn]* [Online]. Available: http://arxiv.org/abs/1007.0810.
TRINH, K. T. 2010d. A Zonal Similarity Analysis of Velocity Profiles in Wall-Bounded Turbulent Shear Flows. *arXiv.org 1001.1594 [phys.fluid-dyn]* [Online].
YOO, S. S. 1974. *Heat transfer and friction factors for non-Newtonian fluids in turbulent flow, PhD thesis,* US, University of Illinois at Chicago Circle.
ZHU, H., KIM, Y. & DE KEE, D. 2005. Non-Newtonian fluids with a yield stress. *Journal Of Non-Newtonian Fluid Mechanics,* 129**,** 177-181.